\documentclass[aps,pra,reprint,superscriptaddress]{revtex4-2}
\pdfoutput=1
\usepackage{color}
\usepackage[normalem]{ulem}
\usepackage{graphicx}
\usepackage{amsmath}
\usepackage{amsfonts}
\usepackage{amssymb}
\usepackage{microtype}
\usepackage{verbatim}
\usepackage{float}
\usepackage{bbold}
\usepackage{hyperref}
\usepackage{bbm}
\usepackage{braket}
\usepackage{cleveref}

\hypersetup{
  colorlinks   = true, 
  urlcolor     = blue, 
  linkcolor    = blue, 
  citecolor   = blue 
}
\usepackage{textcomp}

\usepackage[smallerops]{newtxmath}
\graphicspath{{./Figures/}}

\def\a{\hat{a}}
\def\S{\hat{S}}
\def\G{\hat{G}}
\def\g{\gamma}

\begin{document}

\author{Agnieszka Wierzchucka}
\email{agnieszka.wierzchucka@merton.ox.ac.uk}
\affiliation{Max Planck Institute for the Physics of Complex Systems, 01187 Dresden, Germany}
\affiliation{Merton College, University of Oxford, Oxford OX1 4JD, U.K.}

\author{Francesco Piazza}
\affiliation{Max Planck Institute for the Physics of Complex Systems, 01187 Dresden, Germany}
\affiliation{Theoretical Physics III, Center for Electronic Correlations and Magnetism, Institute of Physics, University of Augsburg, 86135 Augsburg, Germany}

\author{Pieter W. Claeys}
\affiliation{Max Planck Institute for the Physics of Complex Systems, 01187 Dresden, Germany}

\title{Integrability, multifractality, and two-photon dynamics in disordered Tavis-Cummings models}

\begin{abstract}
The Tavis-Cummings model is a paradigmatic central-mode model where a set of two-level quantum emitters (spins) are coupled to a collective cavity mode.
Here we study the eigenstate spectrum, its localization properties and the effect on dynamics, focusing on the two-excitation sector relevant for nonlinear photonics.
These models admit two sources of disorder: in the coupling between the spins and the cavity and in the energy shifts of the individual spins.
While this model was known to be exactly solvable in the limit of a homogeneous coupling and inhomogeneous energy shifts, we here establish the solvability in the opposite limit of a homogeneous energy shift and inhomogeneous coupling, presenting the exact solution and corresponding conserved quantities.
We identify three different classes of eigenstates, exhibiting different degrees of multifractality and semilocalization closely tied to the integrable points, and study their stability to perturbations away from these solvable points. The dynamics of the cavity occupation number away from equilibrium, exhibiting boson bunching and a two-photon blockade, is explicitly related to the localization properties of the eigenstates and illustrates how these models support a collective spin description despite the presence of disorder.
\end{abstract}

\maketitle

\section{Introduction}
In the past years cavity quantum electrodynamics (QED) has emerged as a new platform for the quantum simulation of light-matter interactions~\cite{mivehvar_cavity_2021}. In such cavity setups the electromagnetic cavity mode is coherently coupled to a set of quantum emitters behaving as pseudospins, allowing for polaritons arising as a strong hybridization between the photonic mode of the electromagnetic field and the emitter modes. Furthermore, the cavity-mediated interactions in these models are highly tunable, allowing for systematic studies of e.g. the role of disorder in a controlled way~\cite{lewis-swan_cavity-qed_2021,sauerwein_engineering_2023}. 
Motivated by these experimental developments, there have been a wealth of studies on the interplay between strong light-matter coupling and disorder, specifically on the consequences on the localization properties of the eigenstates~\cite{botzung_dark_2020,dubail_large_2022,engelhardt_polariton_2023,mattiotti_multifractality_2023}. Cavity models can exhibit `bright' polaritonic eigenstates with strong light-matter hybridization and `dark' states with minimal entanglement between the cavity  and the emitters, and different classes of states can have different localization propreties that respond differently to the presence of disorder, which is in turn reflected in the dynamical properties of these models~\cite{fink_dressed_2009}. Ref.~\cite{botzung_dark_2020} argued that dark states exhibit `semilocalization', being localized on multiple noncontiguous sites. Bright and dark states were also shown to exhibit multifractality, being neither fully localized nor delocalized, in a single-excitation setup described by arrowhead matrices~\cite{dubail_large_2022}.

However, the presence of disorder is not the only factor that needs to be taken into account: the many-body dynamics and eigenstate properties can be modified by integrability. Various cavity models are either integrable or close to an integrable point, where the model can be exactly solved using Bethe ansatz and the dynamics is constrained by an extensive amount of conservation laws. 
Variants of the Tavis-Cummings model, the main focus of this work, were originally shown to be integrable by Gaudin~\cite{gaudin_bethe_2014} and later a set of conserved charges were identified by Dukelsky \emph{et al.}~\cite{dukelsky_exactly-solvable_2004}. Such cavity models exhibiting `one-to-all' interactions are part of a family of integrable Richardson-Gaudin Hamiltonians~\cite{dukelsky_colloquium_2004,ortiz_exactly-solvable_2005,claeys_richardson-gaudin_2018}. Crucially, these models remain integrable in the presence of disorder, and their eigenstates have been shown to exhibit anomalous localization properties~\cite{buccheri_structure_2011}. 
Indeed, the effect of integrability was subsequently studied in Ref.~\cite{mattiotti_multifractality_2023}, where it was shown that integrability-breaking restores the ergodicity of the eigenstates in the thermodynamic limit for a finite density of excitations.

One paradigmatic model in cavity QED is the Tavis-Cummings model \cite{tavis_exact_1968,grynberg_introduction_2010}, describing the interaction of $N$ pseudospins with a central bosonic mode under the rotating-wave approximation \cite{jaynes_comparison_1963}. Its Hamiltonian can be written as
\begin{align}\label{eq:H_TC}
\hat{H}_{\textrm{TC}} =    \Delta\, \a^{\dagger} \a + \sum_{i = 1}^{N} \epsilon_i \left(\S_i^z+\frac{1}{2}\right) + \frac{1}{\sqrt{N}} \sum_{i = 1}^{N} \g_i\left(\S_i^{+} \a + \S_i^{-}\a^{\dagger}\right),
\end{align}
where $\a^{\dagger} (\a)$ are the bosonic creation (annihilation) operators for the cavity mode and $\S_i^{\alpha}$ spin operators describing pseudospin $i$. We will assume that these are spin-$1/2$ pseudospins, representing two-level quantum emitters, but this is not a necessary assumption. This model can be realized in both cavity and circuit QED~\cite{wallraff_strong_2004,fink_climbing_2008,fink_dressed_2009,sauerwein_engineering_2023}. The first term gives the energy of the bosonic mode and the second term that of the quantum emitters (each with bare Zeeman energy $\epsilon_i$). The final term describes the inhomogeneous coupling between the bosonic mode and the pseudospins with interactions strengths $\g_i$. Here we have allowed for disorder in both the bare energies and the couplings between the cavity mode and the atoms. The Hamiltonian commutes with the total excitation number, $M = \hat{a}^{\dagger}\hat{a} +  \sum_{i = 1}^{N} \S_i^{+} \S_i^{-}$, such that we can restrict ourselves to sectors with a fixed number of excitations. For any sector with a finite number of excitations, i.e. the number of excitations does not scale with $N$, the factor $1/\sqrt{N}$ in the interaction term is required to have an finite energy in the thermodynamic limit of infinite system size $N \to \infty$. In this way this factor plays the same role as the Kac factor in systems with long-range interactions \cite{kac_critical_1969}, an analogy that will be elaborated in the remainder of this work.

Theoretical studies of the model in the one excitation sector, $M=1$, have already been carried out in Ref.~\cite{botzung_dark_2020}, illustrating the multi-fractal structure of the eigenstates along with their semilocalization properties. The dynamics in this sector was also studied in Ref.~\cite{sun_dynamics_2022}, where the presence of disorder gave rise to a variety of complex behaviors. However, nonlinear photonics require extending the Hilbert space to include multiple excitation~\cite{chang_quantum_2014}. Here, we work within the two excitation sector, $M=2$, to study the localization properties of the eigenstates and their effect on the dynamics. While the model has long been known to be integrable for homogeneous coupling strengths and inhomogeneous  bare energies, we show that the opposite limit is also integrable. This exact solution is remarkably transparent and allows us to identify three different classes of eigenstates. These two integrable limits conspire to result in multifractal eigenstates where the localization properties are surprisingly robust to integrability-breaking perturbations. 

The eigenstates properties are directly reflected in the dynamics, and we use the nearby integrable limit to present exact results for the short-time dynamics of disordered Tavis-Cummings models. We note that the structure of these cavity models is closely related to central spin models, where a set of noninteracting spins interact with a central spin, similar to the cavity mode in these setups. Such central spin models similarly support both bright and dark states~\cite{taylor2003long,taylor_controlling_2003,imamoglu_optical_2003,kurucz2009qubit,belthangady2013dressed,villazon_integrability_2020,Wu2020,dimo_strong-coupling_2022,van_tonder_supersymmetry_2023}, now depending on the hybridization between the central spin and the environment, and recent studies have shown that these also exhibit the combination of Richardson-Gaudin integrability and semilocalization and multifractality~\cite{tang_integrability_2023}.

This work is structured as follows. In Sec.~\ref{sec:homogeneous_TC} we consider the Tavis-Cummings Hamiltonian for two excitations in the absence of disorder, highlighting the existence of three classes of eigenstates. In Sec.~\ref{sec:disorder_g} we present the exact solution for the model in the presence of disorder in the interaction strenghts, reducing the diagonalization of the Hamiltonian to solving a single nonlinear equation with different classes of solutions, and present the exact conserved charges. For interaction strengths scaling as in Eq.~\eqref{eq:H_TC}, we show how the dynamics are further constrained due to the approximate conservation of permutation operators and derive the corresponding relaxation times in Sec.~\ref{sec:relaxation}. For completeness, the exact solution of the model for inhomogeneous bare energies is presented in Sec.~\ref{sec:disorder_eps}. The localization properties of the three different classes of 
eigenstates are discussed in Sec.~\ref{sec:IPR}, and these are related to different examples of the dynamics of the cavity mode occupation number in Sec.~\ref{sec:dynamics}. Sec.~\ref{sec:conclusion} presents our conclusions. 

\section{The homogeneous Tavis-Cummings Hamiltonian}
\label{sec:homogeneous_TC}

In order to understand the spectrum of the two-excitation Tavis-Cummings Hamiltonian and distinguish different classes of eigenstates, it is instructive to first consider the homogeneous limit. In the absence of disorder, when all the bare energies and coupling strengths are equal, the Tavis-Cummings Hamiltonian can be solved by introducing collective spin operators $\S^{\alpha}_{\rm tot} = \sum_{i=1}^N \S_i^{\alpha}$. The Hamiltonian~\eqref{eq:H_TC} reduces to
\begin{align}
\hat{H}_{\textrm{TC}} =  \Delta \,\a^{\dagger} \a + \epsilon \left(\S^z_{\rm tot}+\frac{N}{2}\right) +  \frac{\g}{\sqrt{N}} \left( \a\, \S^{+}_{\rm tot}+ \a^{\dagger}\S^{-}_{\rm tot}\right) ,
\end{align}
where we have set $\epsilon_i = \epsilon, \forall i$ and $\g_i = \g, \forall i$. The Hamiltonian can now be expressed by introducing total spin states $\ket{S_{\rm tot}, S^z_{\rm tot}} = \ket{S,M_S}$, with total spin $S$ and total spin projection $M_S$. 

The restriction to the two-excitation sector requires that there can only be up to two spin excitations, such that $M_S$ can only take the values $-N/2$, $-N/2+1$ and $N/2+2$ (in what follows, we assume that $N\geq 4$). The Hamiltonian additionally conserves total spin quantum number, which can take a maximal value of $S=N/2$, such that the Hamiltonian decomposes in three blocks with different total spin. Either $S=N/2-2$, for which there is a single possible state
\begin{align}\label{eq:singletbasis}
\ket{0} \otimes \ket{N/2-2,-N/2+2},
\end{align}
or $S=N/2-1$, resulting in two states
\begin{align}\label{eq:doubletbasis}
&\ket{0} \otimes \ket{N/2-1,-N/2+2}, \nonumber\\
&\ket{1} \otimes \ket{N/2-1,-N/2+1},
\end{align}
or $S=N/2$, resulting in three states
\begin{align}\label{eq:tripletbasis}
&\ket{0} \otimes \ket{N/2,-N/2+2},\nonumber \\
&\ket{1} \otimes \ket{N/2,-N/2+1}, \nonumber\\
&\ket{2} \otimes \ket{N/2,-N/2}.
\end{align}
\begin{figure}[t!]                      
 \begin{center}
 \includegraphics[width=\columnwidth]{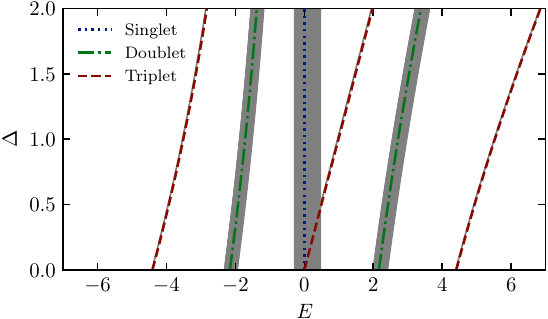}
 \caption{Eigenspectrum of the clean and disordered Tavis- Cummings Hamiltonian for two excitations. In the absence of disorder three classes of eigenstates can be observed resulting in dark states (singlet states) and polaritons (doublet and triplet states). For disordered systems (gray lines) the dark states and the doublet polaritons split into bands of states, while the triplet polaritons remain isolated. Parameters: $N=40$, $\epsilon_i \in [-0.1,0.1]$ and $\g_i \in [1,2]$ uniformly distributed for the disordered model and $\epsilon = \overline{\epsilon_i}$ and $\g^2 = \overline{\g_i^2}$ for the homogeneous model.
 \label{fig:spectrum}}
 \end{center}
\vspace{-\baselineskip}
\end{figure}
In this basis of multiplet states the Hamiltonian takes the form
\begin{align}\label{eq:TC_homogeneous}
\hat{H} = 
\left(
\begin{array}{c|cc|ccc}
2 \epsilon & 0 & 0 & 0 & 0 & 0 \\ \hline
0 & 2 \epsilon & \g & 0 & 0 & 0 \\
0 & \g & \epsilon+\Delta & 0 & 0 & 0 \\ \hline
0 & 0 & 0 & 2 \epsilon & \sqrt{2}\g & 0 \\
0 & 0 & 0 & \sqrt{2}\g & \epsilon+\Delta & \sqrt{2}\g \\
0 & 0 & 0 & 0 & \sqrt{2}\g & 2\Delta \\
\end{array}
\right).
\end{align}
Here we have taken the limit of $N \gg 1$ in all matrix elements and neglected subleading corrections in $1/N$, which however do not change the structure of this matrix.

This Hamiltonian clearly returns three classes of eigenstates, corresponding to the three different blocks: the singlet states from Eq.~\eqref{eq:singletbasis} return eigenstates $\ket{0} \otimes \ket{N/2-2,-N/2+2}$ where the spin degrees of freedom do not hybridize with the cavity mode. These states act as \emph{dark states}. The doublet states from Eq.~\eqref{eq:doubletbasis} hybridize states with no photon in the cavity and with a single photon in the cavity. We will refer to these states as \emph{doublet polaritons}. Finally, the triplet states \eqref{eq:tripletbasis} result in states that are a linear combination of zero, one, or two photonic excitations, such that we will refer to these as \emph{triplet polaritons}. While in the literature the term polaritons is usually reserved for the strong-coupling regime, we will always refer to these states as polaritons for clarity (also following Ref.~\cite{dubail_large_2022}). 

The degeneracy of these eigenvalues is given by the total number of ways in which $N$ spin-$1/2$ particles can be coupled to total spin $S$. This number of ways follows from Catalan's triangle, resulting in a total degeneracy of $N(N-3)/2$ for the singlet state, a total degeneracy of $(N-1)$ for each of the doublet states, and the triplet states are nondegenerate since there is only a single way of coupling $N$ spin-$1/2$ particles to total spin $S=N/2$.

\section{Disordered couplings}
\label{sec:disorder_g}

The eigenspectrum of the Tavis-Cummings model is illustrated in Fig.~\ref{fig:spectrum} for varying $\Delta$, both in the absence and presence of disorder. In the presence of disorder the highly degenerate dark states and the doublet polaritons split into bands of states, with the number of states corresponding to the degeneracy in the absence of disorder, whereas the triplet polaritons remain isolated states. For sufficiently weak disorder strengths the different bands do not overlap, which we will take to be the case in the remainder of this work. 

In the following we will show that the presence of weak disorder in the coupling between the cavity mode and the pseudospins indeed preserves the three classes of eigenstates. In this limit the Hamiltonian can be written as
\begin{align}\label{eq:H_TC_inhomo_g}
\hat{H}_{\textrm{TC}} =  \Delta\, \a^{\dagger} \a + \epsilon \sum_{i = 1}^{N} \left(\S_i^z+\frac{1}{2}\right) + \left(\a\, \G^+  + \a^{\dagger}\G^-\right),
\end{align}
with
\begin{align}
\G^{\pm} = \frac{1}{\sqrt{N}}\sum_{i=1}^N \g_i \S_i^{\pm} = \sum_{i=1}^N g_i \S_i^{\pm}\,.
\end{align}
For convenience we have defined $g_i = \gamma_i / \sqrt{N}$. 
Since the Hamiltonian commutes with the total excitation number $\hat{M} = \hat{a}^{\dagger}\hat{a} +  \sum_{i = 1}^{N} \S_i^{+} \S_i^{-}$, we can set $\epsilon$ to zero without loss of generality.

We first present an explicit construction of the eigenstates, proving the exact solvability of the Tavis-Cummings model (in the two-excitation sector) for disordered couplings and in the absence of disorder in the bare energies. This exact solution then also allows us to construct an extensive set of conserved charges commuting with the Hamiltonian. The integrability of this model was conjectured in Ref.~\cite{tang_integrability_2023}, and we here prove a limited version of this conjecture, showing that the integrability holds if the model is restricted to the two-excitation sector.

\subsection*{Dark states}
In the presence of a disordered coupling the model still allows exact dark states. These are eigenstates of the Hamiltonian \eqref{eq:H_TC_inhomo_g} of the form $\ket{0}\otimes \ket{\mathcal{D}}$, where the spin state satisfies
\begin{align}
\G^-\ket{\mathcal{D}} = 0\,.
\end{align}
These states are adiabatically connected to the singlet states from Eq.~\eqref{eq:singletbasis}, since in the limit of a homogeneous interaction strength $\G^- \propto \S^-_{\textrm{tot}}$, and are well studied in the literature on central mode models~\cite{taylor2003long,taylor_controlling_2003,imamoglu_optical_2003,kurucz2009qubit,belthangady2013dressed,villazon_integrability_2020,Wu2020,dimo_strong-coupling_2022,van_tonder_supersymmetry_2023}. The total number of dark states is $N(N-3)/2$, and these states are annihilated by the interaction part in the Hamiltonian, such that these are again (highly degenerate) eigenstates of the Hamiltonian \eqref{eq:H_TC_inhomo_g} with eigenvalue $2 \epsilon$, since by construction
\begin{align}
\left[\Delta\, \a^{\dagger} \a + \left(\a\, \G^+  + \a^{\dagger}\G^-\right)\right] \ket{0}\otimes \ket{\mathcal{D}} = 0\,.
\end{align}

\begin{figure}[t!]                      
 \begin{center}
 \includegraphics[width=\columnwidth]{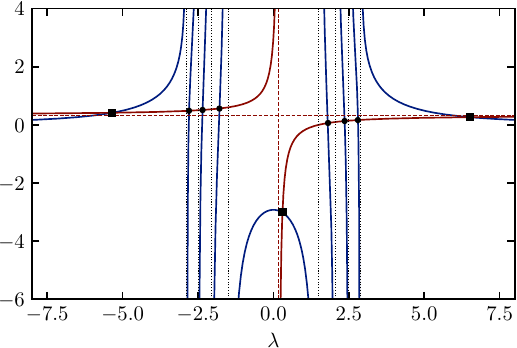}
 \caption{Graphical illustration of the secular equation. Each intersection between the left-hand side (red line) and right-hand side (blue line) returns the eigenvalue of a polariton state, leading to two classes of solutions: doublet polaritons (circles) and triplet polaritons (squares). Parameters: $\Delta=1$, $\epsilon=0$, $N=4$ and $g_i^2 = 1,\dots, 4.$
 \label{fig:secular}}
 \end{center}
\vspace{-\baselineskip}
\end{figure}

\subsection*{Polaritons}

For the polaritons, the exact diagonalization of the Hamiltonian \eqref{eq:H_TC_inhomo_g} can be reduced to solving a nonlinear equation for the eigenvalues. In the spirit of integrability, we can consider an ansatz for the polariton eigenstates with a single free parameter, in such a way that the eigenvalue equation reduces to a nonlinear equation for this parameter.

Specifically, we consider an ansatz for the (unnormalized) eigenstates of the form
\begin{align}\label{eq:bright_ansatz}
\ket{\mathcal{B}(\kappa)} = \left[1+\frac{\a\, \G^+}{\kappa + \Delta} +\frac{\a^{\dagger} \G^-}{\kappa - \Delta}\right]\ket{1} \otimes \ket{\phi_{\kappa}},
\end{align}
where $\kappa$ is a free variable and $\ket{\phi_{\kappa}}$ is a wave function for the $N$ spin degrees of freedom with a single spin excitation, dependent on $\kappa$. Acting with the Hamiltonian~\eqref{eq:H_TC_inhomo_g} on the ansatz \eqref{eq:bright_ansatz}, it is straightforward to show that this state is an eigenstate with eigenvalue $\kappa + \Delta$ provided
\begin{align}\label{eq:selfconsistent}
 \hat{H}(\kappa) \ket{\phi_{\kappa}} \equiv  \left[ \frac{\G^- \G^+}{\kappa + \Delta} + 2\frac{\G^+ \G^-}{\kappa - \Delta} \right]\ket{\phi_{\kappa}}= \kappa \ket{\phi_{\kappa}}. 
\end{align}
This equation is a self-consistent eigenvalue equation since the spin Hamiltonian $\hat{H}(\kappa)$ explicitly depends on the eigenvalue $\kappa$. 
Note also that for a normalized state $\ket{\phi_{\kappa}}$ the norm of the polariton wave functions~\eqref{eq:bright_ansatz} can be calculated from the Hellmann-Feynman theorem as 
\begin{align}
\braket{\mathcal{B}(\kappa)|\mathcal{B}(\kappa)} &= 1+ \frac{\braket{\phi_{\kappa}|\G^-\G^+|\phi_{\kappa}}}{(\kappa+\Delta)^2} +2 \frac{\braket{\phi_{\kappa}|\G^+\G^-|\phi_{\kappa}}}{(\kappa-\Delta)^2} \nonumber\\
&= 1 - \partial_{\kappa}\langle \hat{H}(\kappa) \rangle\,.
\end{align}

Crucially, the Hamiltonian in Eq.~\eqref{eq:selfconsistent} is integrable for every choice of $\kappa$ and its eigenvalues and eigenstates can be explicitly constructed. 
As shown in Appendix~\ref{app:Bethe_derivation}, using this exact solution the self-consistent eigenvalue equation can be recast as a secular equation. Defining $\bar{g}^2 = \sum_{i = 1}^N g_i^2$, $\tilde{\Delta} = \Delta/2$, a state $\ket{\phi_{\kappa}}$ can be written as
\begin{align}\label{eq:phi_kappa}
\ket{\phi_{\kappa}} = \left( \sum_{i = 1}^N \frac{g_i}{\lambda^2 - \tilde{\Delta}^2 -\bar{g}^2+ 2g_i^2 } \S^+_i \right)\ket{\emptyset}\,,
\end{align}
with $\ket{\emptyset} = \ket{\downarrow, \dots ,\downarrow}$. The eigenvalue equation for the polariton energy $E$ then reduces to solving the following equation for $\lambda = E-\tilde{\Delta} = \kappa+\tilde{\Delta}$:
\begin{equation}\label{eq:secular}
    \frac{\lambda - 3\tilde{\Delta}}{3\lambda - \tilde{\Delta}} = \sum_{i = 1}^{N}\frac{g_i^2}{ \lambda^2 - \tilde{\Delta}^2 -\bar{g}^2+ 2g_i^2}\,.
\end{equation}

Remarkably, different solutions to this equation can be directly identified with either the doublet or triplet polaritons. In Fig.~\ref{fig:secular} the structure of this equation is made explicit by plotting both sides as a function of $\lambda$, and the intersections between these curves correspond to the solutions of Eq.~\eqref{eq:secular}. 

The left-hand side has a vertical asymptote at $\lambda = \tilde{\Delta}/3$ and a horizontal asymptote at $1/3$, and is monotonically increasing everywhere.
The right-hand side is an even function of $\lambda$ and has a series of poles at
\begin{align}\label{eq:poles}
\lambda = \pm \sqrt{\tilde{\Delta}^2+\overline{g}^2-2 g_i^2}\,, \quad i=1\dots N,
\end{align}
with each pair of poles corresponding to a coupling strength $g_i$. The function is monotonically decreasing (increasing) for $\lambda$ positive (negative) and goes to zero for $|\lambda| \to \infty$. There are two classes of solutions to these equations: the solutions corresponding to the doublet states lie in between two poles, leading to two sets of $(N-1)$ states. The three triplet states correspond to the remaining isolated solutions to these equations away from the poles. Taking these together, we find the two curves generally have $2N+1$ intersections, exhausting the available polariton Hilbert space. 
Note that for a sufficiently asymmetric distribution of interaction strenghts a pair of poles can vanish, but the secular equation will still have $2N+1$ solutions, as discussed in Appendix~\ref{app:vanishing_pole}.

We emphasize that, while diagonalizing the exact Hamiltonian rapidly becomes unfeasible for large system sizes, the secular equation \eqref{eq:secular} can be efficiently solved for arbitrarily large system sizes using e.g. an intersection method, where specific triplet or doublet states can be systematically targeted since their relative position w.r.t. the poles is known.

In the limit where $\lambda$ is far away from the poles, the secular equation approximately reduces to a cubic equation
\begin{align}
    \frac{\lambda - 3\tilde{\Delta}}{3\lambda - \tilde{\Delta}} \approx \frac{\bar{g}^2}{\lambda^2 - \tilde{\Delta}^2 -\bar{g}^2  },
\end{align}
with three solutions and corresponding energies $E = \lambda + \bar{\Delta}$ given by:
\begin{equation}
    \label{eq:cubicsolutions}
    E = 
    \begin{cases} 
    2\tilde{\Delta} -2\sqrt{\tilde{\Delta}^2 + \bar{g}^2} & \\
    2\tilde{\Delta} & \\
    2\tilde{\Delta} +2\sqrt{\tilde{\Delta}^2 + \bar{g}^2} & \\
    \end{cases}
\end{equation}
These eigenvalues now reduce to the eigenvalues for the triplet states from Eq.~\eqref{eq:TC_homogeneous}.

We note that the secular equation \eqref{eq:secular} closely resembles the dispersion equation within the random phase approximation (RPA), which aims to construct approximate particle-hole excitations on top of a reference state~\cite{rowe_nuclear_1970}. This similarity arises more generally within the theory of Richardson-Gaudin integrability (see e.g. Ref.~\cite{de_baerdemacker_richardson-gaudin_2012}): for single-excitations states the Bethe equations typically reduce to the dispersion relation for the Tamm-Dancoff approximation (TDA), which aims to construct particle-like excitations on top of a reference state, and the RPA can be seen as the two-excitation generalization of the TDA~\cite{rowe_nuclear_1970}.

\subsection*{Conservation laws}
The conserved charges for the Hamiltonian \eqref{eq:H_TC_inhomo_g} can be constructed from those of the factorizable Richardson-Gaudin Hamiltonians~\cite{gaudin_bethe_2014,dukelsky_colloquium_2004,claeys_richardson-gaudin_2018}. These conserved charges are easiest to represent in a block matrix representation, similar to Eq.~\eqref{eq:TC_homogeneous}, and are of the form
\begin{equation}
    Q_j = \begin{pmatrix}
        -\frac{2}{3}\Delta( \tilde{Q}_j + S_j^z) & G^+(\tilde{Q}_j + S^z_j/3) & 0 \\
        (\tilde{Q}_j + S_j^z/3)G^- & \frac{\Delta}{3}(\tilde{Q}_j + 3 S_j^z) & \sqrt{2}(\tilde{Q}_j + S_j^z/3)G^+ \\
        0 & \sqrt{2}(\tilde{Q}_j + S_j^z/3) & \frac{4}{3}\Delta(\tilde{Q}_j - S_j^z)
    \end{pmatrix},
\end{equation}
where $\tilde{Q}_j$ are operators acting on the spin degrees of freedom as
\begin{align}\label{eq:factorizablecharges}
    \tilde{Q}_j &=\frac{S_j^+S_j^- + S_j^-S_j^+}{2} \nonumber \\ &+ \sum_{k \neq j}^N \left(\frac{g_j g_k}{g_j^2 - g_k^2}(S_j^+ S_k^- +  S_j^- S_k^+) + \frac{2g_k^2}{g_j^2 - g_k^2} S_j^z S_k^z \right)\,.
\end{align}
That they commute with the Hamiltonian \eqref{eq:H_TC_inhomo_g} for two excitations can be verified by direct calculation and is shown in Appendix~\ref{app:charges}. The existence of these conserved quantities guarantees the integrability of the model. We note that this calculation is highly similar to a recent calculation of the conserved charges in a spin-$1$ central spin model~\cite{tang_integrability_2023}, a related central mode model where a similar block matrix structure appears. For this reason, we will defer from discussings these conserved charges in more detail and refer the reader to Ref.~\cite{tang_integrability_2023}. We only note that this previous work conjectured the integrability of the Tavis-Cummings model with disordered couplings (for an arbitrary number of excitations), and this work establishes its integrability in the limiting case of two excitations. The integrability for a higher number of excitations remains an open question.

\section{Relaxation times for permutation symmetry}
\label{sec:relaxation}

The Bethe states in the previous derivation can be adiabatically connected to collective spin states. 
This correspondence suggests that, at least for not too strong disorder, the considered model can be described in terms of the collective spin operators of the homogeneous model. 
Here we show that for the Hamiltonian~\eqref{eq:H_TC} such a description of the dynamics is justified up to times scaling as $t \propto \sqrt{N}$ in the absence of disorder in the bare energies. 
More specifically, collective spin states are indicative of an underlying spin permutation symmetry, and we show that this permutation symmetry is preserved up to relaxation times scaling as $\sqrt{N}$ in the new integrable model. This derivation builds on a similar argument for systems with sufficiently long-range interactions~\cite{mori_prethermalization_2019}. Such models similarly support a description in terms of collective spin states in the presence of a nearby integrable (but quasiclassical) limit~\cite{mori_prethermalization_2019,defenu_out--equilibrium_2023,defenu_long-range_2023,lerose_theory_2023}. These relaxation times then present an additional similarity between the current model and lattice systems with sufficiently long-range interactions. In such systems the presence of an integrable semiclassical limit was recently argued to be crucial~\cite{lerose_theory_2023} for such a description, and we here show the stability of the collective spin description near the integrable point.

In the homogeneous models all spin modes are identical, i.e. the Hamiltonian is invariant under any permutation of the spins. The Hamiltonian commutes with permutation operators $\hat{P}_{ij}$, permuting spins $i$ and $j$, and e.g. for spin-$1/2$ we can write these permutations in terms of Pauli matrices as
\begin{align}
\hat{P}_{ij} = \frac{1}{2}\left(\mathbbm{1}_{ij}+\vec{\sigma}_i \cdot \vec{\sigma}_j\right)\,.
\end{align}
These operators are exactly conserved in the dynamics of the homogeneous model, for which the eigenstates in terms of collective spins are similarly eigenstates of the permutation operators. While the permutations operators are no longer exactly conserved in the inhomogeneous case, it is possible to derive the inequality
\begin{align}\label{eq:tau_ij}
|\braket{\hat{P}_{ij}(t)}-\braket{\hat{P}_{ij}(0)}| \leq 4 \sqrt{2} \frac{|\g_i-\g_j|t}{\sqrt{N}}, 
\end{align}
for systems with homogeneous bare energies, indicating that permutation symmetry is conserved up to a time scale $\tau_{ij} \propto \sqrt{N}/|\g_i-\g_j|$. As such, for any initial state that satisfied this permutation symmetry, the dynamics up until the time scale $\tau_{ij}$ can be accurately modeled using the collective spin operators, since these are exactly the operators that do not take the states out of the initial symmetry sector. In the case of homogeneous bare energies, $|\g_i-\g_j|$ scales (at worst) as $1/\sqrt{N}$, resulting in time scales $\tau_{ij}\propto \sqrt{N}$, indicating that any such description becomes increasingly accurate with increasing system size. Note that this is a lower bound -- the scaling as $N^{-1/2}$ holds for the extremal values of $\g_i$ and $\g_j$, whereas the typical closest distance scales as $N^{-3/2}$, indicating much longer-lived conservations for the corresponding spin permutation operator. 

The proof is straightforward and follows a similar proof for systems with long-range interactions from Ref.~\cite{mori_prethermalization_2019}. We have that
\begin{align}
\frac{d}{dt}\braket{\hat{P}_{ij}(t)} = -i \braket{\left[\hat{P}_{ij},\hat{H}\right]} = -i \braket{\hat{P}_{ij}(\hat{H}-\hat{P}_{ij}\hat{H}\hat{P}_{ij})},
\end{align}
where we have used that $\hat{P}_{ij}^2 = \mathbbm{1}$. It follows that
\begin{align}\label{eq:dPdt}
|\braket{\hat{P}_{ij}(t)}-\braket{\hat{P}_{ij}(0)}| &\leq t || \hat{P}_{ij}(\hat{H}-\hat{P}_{ij}\hat{H}\hat{P}_{ij}) || \nonumber\\
& \leq t || \hat{H}-\hat{P}_{ij}\hat{H}\hat{P}_{ij}||,
\end{align}
where we have bounded the expectation value in terms of the operator norm and used that $||\hat{P}_{ij}|| = 1$. The operator $\hat{P}_{ij}\hat{H}\hat{P}_{ij}$ is the original Hamiltonian with spins $i$ and $j$ exchanged, such that we find
\begin{align}
\hat{H}-\hat{P}_{ij}\hat{H}\hat{P}_{ij} = \frac{\g_i-\g_j}{\sqrt{N}} \left[\left(\S_i^{+}-\S_j^{+}\right) \a + \left(\S_i^{-}-\S_j^{-}\right)\a^{\dagger}\right]\,.
\end{align}
While the operator norm of $\a$ and $\a^{\dagger}$ is in general unbounded, here we can make use of the restriction to the two-excitation subspace, for which $\braket{\a^{\dagger}a} \leq 2$ such that $|| \a || = ||\a^{\dagger}|| = \sqrt{2}$. The operator norm of the above operator norm is bounded by the sum of the operator norms of each term, resulting in
\begin{align}
|| \hat{H}-\hat{P}_{ij}\hat{H}\hat{P}_{ij} || \leq 4 \sqrt{2}\,\frac{|\g_i-\g_j|}{\sqrt{N}}\,.
\end{align}
Plugging this bound in Eq.~\eqref{eq:dPdt}, we obtain the proposed bound from Eq.~\eqref{eq:tau_ij}.

While the integrability of the model guarantees nonergodicity due to the presence of conservation laws, this result further constrains the dynamics. For any initial state that is fully symmetric in the spins, as is e.g. the case for a state with no spin excitations and 2 photonic excitations, the dynamics can be restricted to the Dicke manifold, i.e. spin states that are fully symmetric under spin exchange, up to a time scale scaling as $\sqrt{N}$. This space is also known as the totally symmetric subspace (TSS), and the role of the TSS in spin dynamics has been the subject of active study \cite{qi_surprises_2023}. Only after this relaxation time scale can the system move out of the TSS, and we find that the TSS is increasingly stable for increasing system sizes. 

Note also that the above relaxation time only depends on the restriction to the two-excitation sector by setting $||\a ||=||\a^{\dagger} ||=\sqrt{2}$. For the $n$-excitation sector, in the above derivation the factor $\sqrt{2}$ only needs to be replaced by $\sqrt{n}$. In any sector where the number of excitations does not scale with system size, we hence expect relaxation times for the permutation symmetry scaling as $\sqrt{N}$. If the number of excitations $M$ scales with $N$, i.e. $M \propto N$, then the Kac factor in the Hamiltonian~\eqref{eq:H_TC_inhomo_g} also needs to be modified to $1/N$, i.e.
\begin{align}
\hat{H}_{\textrm{TC}} =   \Delta\, \a^{\dagger} \a + \epsilon \sum_{i = 1}^{N} \left(\S_i^z+\frac{1}{2}\right) + \frac{1}{N} \sum_{i = 1}^{N} \g_i\left(\S_i^{+} \a + \S_i^{-}\a^{\dagger}\right).
\end{align}
Repeating the derivation above directly results in the bound
\begin{align}
|\braket{\hat{P}_{ij}(t)}-\braket{\hat{P}_{ij}(0)}| \leq 4 \sqrt{M} \frac{|\g_i-\g_j|t}{N} \propto |\g_i-\g_j|\frac{t}{\sqrt{N}}, 
\end{align}
again indicating that permutation symmetry is preserved up to time scales scaling as $\sqrt{N}$. We obtain the general result that permutation symmetry is preserved up to time scales scaling as $\sqrt{N}$, irrespective of the number of the excitations, provided that the Hamiltonian is defined in such a way that energy is finite. This argument fails for a disorder in the bare energies $\epsilon_i$, since then all obtained time scales would be ${O}(1)$. In this limit, however, it appears that integrability again stabilizes the collective spin description.

In the context of long-range systems, we note that the scaling with system size of the corresponding relaxation times follows directly from the Kac factor fixing the extensivity of the energy, similar to how the time scale in our context requires the correct normalization of the interaction strengths.

\section{Disordered bare energies}
\label{sec:disorder_eps}
For completeness, we reiterate the exact solution of the Hamiltonian with homogeneous couplings and disorder in the bare energies,
\begin{align}\label{eq:H_TC_inhomo_eps}
\hat{H}_{\textrm{TC}} =  \Delta\, \a^{\dagger} \a + \sum_{i = 1}^{N} \epsilon_i \left(\S_i^z+\frac{1}{2}\right) + g \sum_{i = 1}^{N} \left(\a \, \S_i^{+}  + \a^{\dagger}\S_i^{-}\right)\,.
\end{align}
It is known that in this limit the model is integrable \cite{dukelsky_exactly-solvable_2004} and can be solved using the Bethe ansatz \cite{babelon_bethe_2007} (see also Refs.~\cite{tsyplyatyev_simplified_2010,tschirhart_algebraic_2014,skrypnyk_twisted_2016,skrypnyk_generalized_2008}). The eigenstates can then be written as Bethe states, characterised by two variables $E_1$ and $E_2$:
\begin{equation}\label{eq:BetheAnsatz_2}
     \ket{\psi(E_1, E_2)} = \S^\dagger(E_1) \S^\dagger(E_2)\ket{\emptyset},
\end{equation}
expressed in terms of generalized raising operators
\begin{align}
\S^\dagger(E_{\alpha}) = a^\dagger - g \sum^N_{i = 1} \frac{\S_i^{+}}{\epsilon_i - E_{\alpha}}\,,
\end{align}
acting on the vacuum state $\ket{\emptyset} = \ket{0}\otimes \ket{\downarrow \dots \downarrow}$. These states are eigenstates of the partially homogeneous Hamiltonian with total energy $E = E_1 + E_2$, provided the variables satisfy the set of Bethe equations \cite{tsyplyatyev_simplified_2010}:
\begin{align}
    \Delta - E_1 + \sum_{i = 1}^N \frac{g^2}{E_1 - \epsilon_i} = \frac{2g^2}{E_1 - E_2}, \\ 
    \Delta - E_2 + \sum_{i = 1}^N \frac{g^2}{E_2 - \epsilon_i} = \frac{2g^2}{E_2 - E_1}\,.
\end{align}
The two variables $E_{1,2}$ are also referred to as 'quasi-energies' due to their role in the energy of the eigenstate they describe.

These equations have been well studied in the literature~\cite{babelon_bethe_2007,faribault_gaudin_2011,strater_nonequilibrum_2012,claeys_eigenvalue-based_2015}. For our results it is relevant that there exist different classes of solutions, depending on the position of the variables $E_{1,2}$ with respect to the poles $\epsilon_i$ in the Bethe equations: either both $E_1$ and $E_2$ are both far away from the poles, or one variable is away from the poles and the other is `trapped' between a pair of poles, or both variables `trapped' between pairs of poles. Following our discussion for the limit of inhomogeneous couplings, these solutions can be identified with triplet polaritons, doublet polaritons, and singlet dark states respectively.

The Hamiltonian again supports an extensive set of conserved quantities, one for each spin in the system, where now
\begin{align}
\hat{Q}_j =& (\Delta-\epsilon_j)\S^z_j-g(\S^+_j \a+ \S^-_j \a^{\dagger}) \nonumber\\
&-2 g^2 \sum_{k \neq j}^N \frac{1}{\epsilon_j-\epsilon_k}\left[\frac{1}{2}(\S^+_j \S^-_k+\S^-_j \S^+_k)+\S^z_j \S^z_k\right]\,.
\end{align}
These form a set of mutually commuting conserved charges satisfying $[\hat{H},\hat{Q}_i] = [\hat{Q}_i,\hat{Q}_j]=0, \forall i,j$.

\section{Semilocalization and multifractality}
\label{sec:IPR}
The different classes of eigenstates do not just differ in that they belong to different bands, but they also have different localization properties. Localization for eigenstates in cavity models have recently gained attention because of their anomalous localization properties, which in turn directly translate to a lack of thermalization for a local perturbation~\cite{mattiotti_multifractality_2023}. For a single excitation dark states eigenstates were argued to be `semilocalized', i.e. being localized on multiple noncontiguous sites, in Ref.~\cite{botzung_dark_2020}. In a follow-up work, it was argued that the polaritons in such a model exhibit multifractality: the eigenstates are extended, i.e. not localized, but yet non-ergodic and not fully delocalized~\cite{dubail_large_2022}. Multifractality was similarly observed in the integrable Tavis-Cummings model with a finite excitation density \cite{mattiotti_multifractality_2023}, and arguably already appeared in earlier studies of the (closely related) integrable Richardson model~\cite{buccheri_structure_2011}. This multifractality was similarly observed in Ref.~\cite{tang_integrability_2023} for an integrable central spin model where the central mode is a spin-1 particle, where the dynamics of the central spin mode can serve as a probe for the multifractality.

In order to probe the multifractality of a state $\ket{\psi}$, we consider the $q$-dependent inverse participation ratio, defined as:
\begin{equation}
    \mathcal{P}(q) = \sum_{n=1}^{D} |\braket{n|\psi}|^{2q},
\end{equation}
where $n$ are product states in the full (photonic and spin) $D$-dimensional Hilbert space. The $q$-dependent IPR quantifies the distribution of the components of an eigenstate in a product state basis, with $q$ acting as the equivalent of the order in the R\'enyi entropies. For a delocalized eigenstate all coefficients scale as $1/\sqrt{D}$, resulting in an IPR scaling with dimension of the Hilbert space as $D^{1-q}$. A change in this scaling as $q$ is varied is a signature of multifractality in the eigenstate~\cite{wegner_IPR_1980,evers_IPR_2000}.

The different terms in this summation can be made explicit by labelling the basis states as $\ket{0} \equiv \ket{2} \otimes \ket{\emptyset}$, $\ket{i} \equiv \ket{1}\otimes \S_i^+\ket{\emptyset}$, and $\ket{i,j} \equiv \ket{0} \otimes \S_i^+ \S_j^+ \ket{\emptyset}$ and introducing a corresponding notation for the eigenstate components. The IPR then reads
\begin{align}
\mathcal{P}(q) = |\psi_0|^{2q} + \sum_{i=1}^N |\psi_i|^{2q} + \sum_{i,j=1}^N |\psi_{i,j}|^{2q}\,.
\end{align}
In the following, we show that it is useful to consider the different contributions to the IPR separately and introduce `restricted' versions of the IPR, where the summations are restricted to the sectors with a fixed number of photons. Specifically, we write
\begin{align}
\mathcal{P}_0(q) &= |\psi_0|^{2q} \quad &\textrm{(2-photon)}\,,  \\
\mathcal{P}_1(q) &= \sum_{i=1}^N |\psi_i|^{2q} \quad &\textrm{(1-photon)}\,, \\
\mathcal{P}_2(q) &= \sum_{i,j=1}^N |\psi_{i,j}|^{2q} \quad &\textrm{(0-photon)}\,.
\end{align}
The scaling of the full IPR will be determined by the scaling of the largest term of these three.

\begin{figure}[t!]                      
 \begin{center}
 \includegraphics[width=\columnwidth]{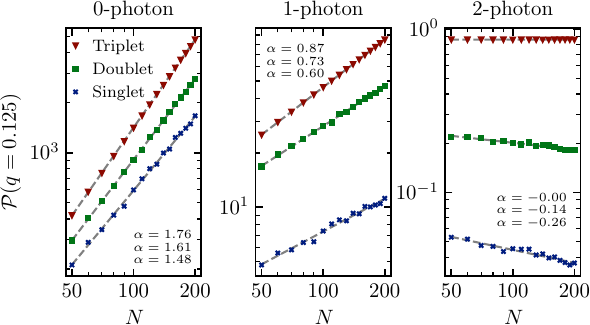}
 \caption{Scaling of the IPR for the three different classes of eigenstates in the different subsection of the Hilbert space for $q=0.125$. Gray dashed lines indicate best fits $\propto N^{\alpha}$. Parameters: $\Delta = 1$,  $\epsilon_i$ uniformly distributed in $[-0.1,0.1]$ and $g_i$ uniformly distributed in $[1,3]/\sqrt{N}$.
 \label{fig:IPR}}
 \end{center}
\vspace{-\baselineskip}
\end{figure}

In Fig.~\ref{fig:IPR} we present numerical result for these three components as a function of system size $N$ in the presence of disorder in the couplings as well as (weak) disorder in the bare energies. The disorder in the bare energies is chosen to be sufficiently weak such that the different bands from Fig.~\ref{fig:spectrum} do not mix, allowing different states to be identified based on their position in the spectrum. 
We consider a single disorder realization, since the fluctuations over the different eigenstates within the bands are smoothened by averaging the IPR within each class of eigenstates.
In this limit we find that all three classes of states exhibit multifractality, with different scaling exponents. 
These scaling exponents are numerically observed to be identical to the exponents obtained in the integrable limit with a homogeneous coupling in all sectors. For inhomogeneous couplings and homogeneous bare energies, care needs to be taken. First of all, the dark states are exactly degenerate, such that it is not meaningful to consider the localization properties of a single dark states. Second, we observe that the scaling exponents in the $0$- and $1$- photon sector are identical to the exponents in the presence of both sources of disorder, but the single amplitude of the $2$-photon component can have different scaling. However, this amplitude does not contribute to the full IPR, such that the scaling of the total IPR will be identical in both the two integrable limits and in the non-integrable case. The scaling of the IPR for the Bethe states is derived in Appendix~\ref{app:IPR_from_Bethe}, and we here focus on the scaling away from these integrable limits. 
\begin{figure*}[ht!]
  \centering \includegraphics[width=0.8\textwidth]{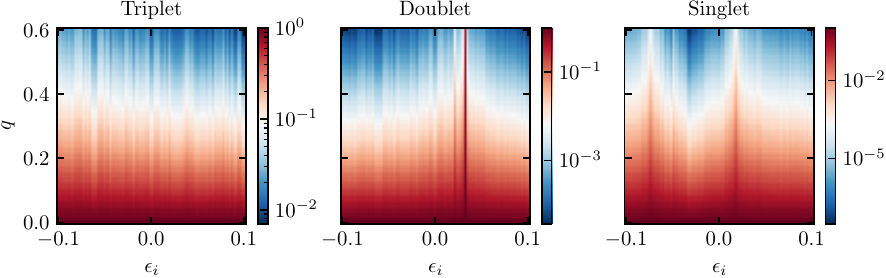}
  \caption{Illustration of the one-photon wave function components $|\psi_i|^{2q}$ for representative polariton (triplet and doublet) and dark (singlet) states. For homogeneous couplings these components are a smooth function of the corresponding bare energy $\epsilon_i$, with different states characterized by the number of poles where the amplitudes scale as $|\psi_i|^{2q} \propto 1/(\epsilon_i-E_{\alpha})^{2q}$. In the presence of disorder in the couplings this overall structure is preserved and in turn reflected in the IPR. Parameters: $N=100$, $\Delta = 1$,  $\epsilon_i$ uniformly distributed in $[-0.1,0.1]$ and $g_i$ uniformly distributed in $[1,2]/\sqrt{N}$.
  \label{fig:multifractality}}
\end{figure*}
\emph{Triplet polaritons.} First focusing on the three triplet polariton states, we observe the following the scaling in the large-$N$ limit:
\begin{align}
\mathcal{P}_0(q) &= O(N^{2-2q}), \label{eq:triplet_P0}\\
\mathcal{P}_1(q) &= O(N^{1-q}), \label{eq:triplet_P1}\\
\mathcal{P}_2(q) &= O(1)\,,	\label{eq:triplet_P2}
\end{align}
independent of $q$. However, the different scaling of these contributions indicate that the total IPR will exhibit a change in scaling as $q$ in varied, with either the $0$-photon term being dominant ($q<1$) or the $2$-photon term being dominant ($q>1$). The IPR for the triplet polaritons will then scale as
\begin{align}\label{eq:IPR_triplet}
    \mathcal{P} (q) & = 
    \begin{cases} 
    O(N^{2-2q}), & \quad 0 < q < 1, \\
    O(1), & \quad q > 1.
    \end{cases} 
\end{align}
This result has a direct interpretation: the total weight of the triplet states within each photon sector is $O(1)$, i.e. the probability of observing $n$ photons in these eigenstates is $O(1)$ for all values of $n$, as quantified in the restricted IPR's for $q=1$:
\begin{align}
\mathcal{P}_{0}(q=1)=\mathcal{P}_{1}(q=1)=\mathcal{P}_{2}(q=1)=O(1),
\end{align}
but within each sector these states are fully delocalized. Recall that a delocalized eigenstate in a Hilbert space of dimension $D$ results in an IPR scaling as $D^{1-q}$ for all values of $q$. The $1$-photon states span a Hilbert space of dimension $N$, and the delocalization in this Hilbert space leads to a scaling $N^{(1-q)}$ in Eq.~\eqref{eq:triplet_P1}. The $0$-photon states span a Hilbert space of dimension $N(N-1)/2 = O(N^2)$ and the delocalization in this space leads to the scaling $N^{2(1-q)}$ in Eq.~\eqref{eq:triplet_P0}. The single $2$-photon state has a weight $O(1)$, leading to the observed scaling from Eq.~\eqref{eq:triplet_P2}. 
While the states are delocalized within the separate photon sectors, they are not delocalized within the full Hilbert space due to the total $O(1)$ weight of the states within each sector. For full delocalization within the Hilbert space the states within each photon sector would not be normalized [up to an $O(1)$ factor] and the restricted IPRs would include additional scaling factors. Delocalization in the full Hilbert space would e.g. predict that the total weight $|\psi_0|^{2}$ of the $2$-photon sector vanishes as the relative dimension of this sector, i.e. $O(N^{-2})$, to be contrasted with the observed $O(1)$ scaling. As such, it is the relative weights of the different sectors that give rise to the change in scaling of the IPR as $q$ is varied, indicating multifractality.

\emph{Doublet polaritons}. For the doublet polaritons, we find that
\begin{align}
	\mathcal{P}_2(q) & =N^{-q} \\
    \mathcal{P}_1 (q) & = 
    \begin{cases} 
    O(N^{1-2q}), & \quad q < 1/2 \\
    O(1), & \quad q \geq 1/2  
    \end{cases} 
    \\ 
    \mathcal{P}_0 (q) & = 
    \begin{cases} 
    O(N^{2-3q}), & \quad q < 1/2 \\  
    O(N^{1-q}), & \quad q \geq 1/2  
    \end{cases}
\end{align}
Remarkably, even within the subsections of the Hilbert space with a fixed number of photons, the eigenstates exhibit multifractality and are not fully delocalized. These scalings reflect underlying semilocalization in the $1$-photon sector, i.e. there is an $O(1)$ number of components dominating the scaling for $q>1/2$, whereas in the $0$-photon sector an $O(N)$ number of states dominate. Interestingly, in the $0$-photon sector the eigenstates are hence localized within a vanishing fraction of the Hilbert space [$O(N)$ components in a $O(N^2)$ Hilbert space]. While the states in both the $0$- and $1$-photon sector are distributed over a vanishing fraction over the Hilbert space, within the $1$-photon sector these are localized on non-contiguous sites, whereas in the $0$-photon sector these are distributed over non-contiguous regions in the Hilbert space (see App.~\ref{app:vanishing_pole}).

The IPR for the doublet polaritons reflects the scaling of the dominant components in the restricted IPR, changing from the scaling of the $0$-photon sector to the scaling of the $1$-photon sector as
\begin{align}
    \mathcal{P} (q) & = 
    \begin{cases} 
    O(N^{2-3q}), & \quad 0 < q < 1/2, \\
    O(N^{1-q}), & \quad  1/2 < q < 1, \\
    O(1), & \quad q > 1.
    \end{cases} 
\end{align}

\emph{Singlet dark states.} The dark states similarly exhibit multifractality within each of the three different sectors, with the restricted IPR scaling as
\begin{align}
	\mathcal{P}_2(q) & =O(N^{-2q}) \\
    \mathcal{P}_1 (q) & = 
    \begin{cases} 
    O(N^{1-3q}), & \quad q < 1/2 \\
    O(N^{-q}), & \quad q \geq 1/2  
    \end{cases} 
    \\ 
    \mathcal{P}_0 (q) & = 
    \begin{cases} 
    O(N^{2-4q}), & \quad q < 1/2 \\  
    O(1), & \quad q \geq 1/2  
    \end{cases}
\end{align}
Taking these results together, the scaling of the IPR follows the $0$-photon IPR, as could be expected for dark states:
\begin{align}
    \mathcal{P} (q) & = 
    \begin{cases} 
    O(N^{2-4q}), & \quad 0 < q < 1/2, \\
    O(1), & \quad q > 1/2.
    \end{cases} 
\end{align}
These scalings now reflect semilocalization in both the $0$ and $1$-photon sector, where in both cases $O(1)$ components dominate the IPR scaling for $q$ large enough.

All presented scalings can be clearly numerically observed in Fig.~\ref{fig:IPR}, where the numerically obtained scaling exponents are close to the theoretically predicted values. However, we emphasize that these results are limited to weak disorder and system sizes up to $N=200$. While it is possible that these scalings break down for larger system sizes, there is no indicator of this happening in our numerics.

These scalings are analytically derived for the integrable limit in Appendix~\ref{app:IPR_from_Bethe}. 
In the same way that the different classes of eigenstates can be connected to the relative position of the Bethe root to the poles in the secular equation~\eqref{eq:secular}, these different scalings can be directly related to the Bethe states. This connection is detailed in Appendix~\ref{app:IPR_from_Bethe}. Any Bethe root that lies close to a pole will lead to large contributions in the eigenstate component related to this pole and corresponding semilocalization on the corresponding basis states. Different localization properties are hence expected for states characterized by a different relative position of the Bethe root to the poles.
These different scalings and the connection with the pole structure can be visualized by considering the eigenstate components within e.g. the single-spin excitation basis states. If the set of bare energies is ordered, these components will be a smooth functions of $\epsilon_i$, with a possible divergence at $E_1$ and $E_2$ if these lie in between two poles. This behavior is illustrated in Fig.~\ref{fig:multifractality}, and it is clear that the presence of (weak) integrability-breaking terms does not change the multifractal character of the eigenstates. For the triplet states all components exhibit the same scaling, where either one or two peaks appear for the doublet and singlet states respectively.

\section{Dynamics and Photon Bunching}
\label{sec:dynamics}
These localization properties directly translate to the dynamical behavior of initial states with a fixed photon number. While it is customary to introduce leakage and describe the dynamics in terms of open systems, we here focus on closed system dynamics in order to keep the connection with the previous results. In this sense these dynamics is expected to be reflective of the short-time dynamics of  realistic cavity models with dissipation, i.e. the dynamics on times shorter than the dissipation time scale.
\begin{figure}[t!]                      
 \begin{center}
 \includegraphics[width=\columnwidth]{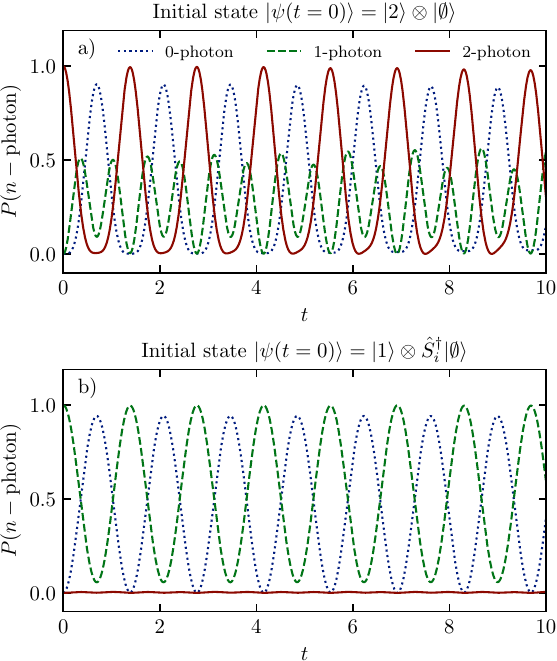}
 \caption{Dynamics of the probability of observing $n$ photons in the cavity for an initial state with $2$ photons (a) and an initial state with $1$ photon and a single cavity excitation (b). Parameters: $N=40$, $g_i$ uniformly distributed in $[1,3]/\sqrt{N}$ and $\epsilon_i$ uniformly distributed in $[-0.2,0.2]$.
 \label{fig:dynamics}}
 \end{center}
\vspace{-\baselineskip}
\end{figure}

Since the restricted IPR indicated different eigenstates localization properties depending on the number of photons, we consider the dynamics of the probability of observing a fixed number of photons in the cavity.
In Fig.~\ref{fig:dynamics} we first consider an initial state with two photons, i.e. $\ket{\psi(t=0)} = \ket{2}\otimes \ket{\emptyset}$. The photon numbers exhibit coherent oscillations with near-perfect revivals and only a slow dephasing. In the large $N$-limit the initial state only has a nonvanishing overlap with the triplet polaritons, following our previous discussion, such that the dynamics can effectively be treated as a three-level system. The period for revivals directly follows from Eq.~\eqref{eq:cubicsolutions} as
\begin{align}
T \approx {2 \pi}/{\sqrt{(\Delta-\overline{\epsilon})^2+\overline{g}^2}}\,.
\end{align}
At integer multiples of the period the system is to good approximation in a 2-photon states, whereas at half-integer multiples the system is close to a 0-photon state, reminiscent of the boson bunching in the Hong-Ou-Mandel effect~\cite{hong_measurement_1987}. These coherent dynamics are now a direct consequence of the multifractality of the triplet polaritons: for delocalized eigenstates all eigenstates would be involved in the dynamics and all coherences would rapidly decay and thermalize, but since the initial (product) state only has $O(1)$ overlap with the triplet polariton the system behaves as a three-level system with long-lived coherences. Following Eq.~\eqref{eq:IPR_triplet} and the surrounding discussion, this $O(1)$ overlap directly relates to the change in IPR as $q$ is varied and hence the multifractality.

\begin{figure}[t!]                      
 \begin{center}
 \includegraphics[width=\columnwidth]{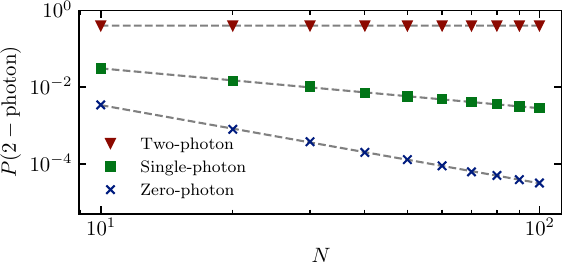}
 \caption{Time-averaged probability of observing $2$ photons in the cavity for an initial state with $n$ photons. Parameters: $\Delta = 1$, $g_i$ equally spaced in $[1,3]/\sqrt{N}$ and $\epsilon_i$ uniformly distributed in $[-0.2,0.2]$.
 \label{fig:twophoton}}
 \end{center}
\vspace{-\baselineskip}
\end{figure}

Second, we consider an initial state with a single photon excitation and a single spin excitation, $\ket{\psi(t=0)} = \ket{1}\otimes \S_i^{\dagger} \ket{\emptyset}$. We first observe that there are again coherent oscillations, now between the $1$-photon and the $0$-photon sector. These oscillations are now a direct consequence of the semilocalization of the doublet states in the $1$-photon sector: the initial state will have $O(1)$ overlap with an $O(1)$ number of doublet polaritons, such that we can again restrict the dynamics to an $O(1)$ number of doublet states. Additionally, the initial state has a vanishing overlap with both the dark states and the triplet polaritons: the former because of the vanishing weight of the dark state in the $1$-photon sector, and the latter because of the delocalization of the triplet polaritons in the one-photon sector, leading to a vanishing overlap between the initial (localized) state and the delocalized eigenstate in this sector. Again, for purely delocalized eigenstates no such coherent dynamics would be observed and the system would rapidly thermalize.

We emphasize that these results conform to the expected cavity dynamics in the homogeneous limit, but here allow for a direct interpretation in terms of the localization properties of the different classes of eigenstates: In the presence of disorder, such coherent oscillations require nontrivial localization properties of the eigenstates within the different $n$-photon sectors. The stability of the eigenstate localization properties to disorder indicates a stability of the dynamics of the homogeneous and integrable limits to the presence of disorder.

The delocalization of the triplet polaritons within the $1$-photon sector can also be directly observed in the vanishing probability of observing $2$ photons in the cavity in Fig.~\ref{fig:dynamics}.b), similar to the two-photon blockade~\cite{kimble_photon_1977,imamoglu_strongly_1997,zubizarreta_casalengua_conventional_2020}. The overlap of the initial state with the triplet polaritons will scale as $O(N^{-1/2})$ due to the delocalization within the $N$-dimensional $1$-photon Hilbert space. Since only these states have a nonvanishing contribution to the probability of observing $2$ photons, the probability of observing a $2$-photon state scales as $O(N^{-1})$. Photon blockade can be attributed to the fact that the states with appreciable two-photon character are delocalized within each photon sector and thus two photons cannot be observed unless they already exist in the initial state (restricting to the reasonable conditions where atoms can be excited only locally).

This argument can be extended to probe the delocalization properties more generally. The time-averaged probability of observing two photons in the cavity for a generic initial state will be due to the overlap of the initial state with the triplet polaritons. For an initial state with $2$ photon excitations we have already argued that this overlap is $O(1)$, such that this time-averaged probability will be $O(1)$. For an initial state with $1$ photon excitation and a single spin excitation this probability scales as $O(N^{-1})$, and for an initial state with $0$ photon excitations and two spin excitations this probability scales as $O(N^{-2})$. These three different scales directly reflect the delocalization within different subspaces, and are illustrated in Fig.~\ref{fig:twophoton}. In order to avoid averaging, we consider a so-called 'picket-fence' model of evenly spaced interaction strengths $g_i$. The different scalings can be clearly observed, relating the different scalings of the restricted IPRs for the triplet polaritons to a physical observable.

\section{Conclusion and discussion}
\label{sec:conclusion}
We considered the Tavis-Cummings model in the presence of disorder in both the bare energies and the interaction strengths, focusing on the sector with two excitations.
In the absence of disorder in the bare energies but for disorder in the interaction strengths, we derived an exact solution for the eigenstates and eigenvalues of the model. The model supports dark states, where the cavity mode does not hybdridize with the spins, and polariton states, where it does so. The dark states are known, and for the polaritons we show how the diagonalization of the full Hamiltonian can be reduced to solving a single secular equation, which can be numerically done in a straightforward way. Every eigenvalue corresponds to a solution of this equation, and we can identify different classes of eigenstates with different localization properties.

The main advances of this work are that we: (i) introduce a new solvable limit of the disordered Tavis-Cummings model, relevant for nonlinear photonics,  presenting its exact solution and conservation laws, (ii) strengthen the connection between integrability and multifractality by analytically showing how the two integrable limits of this model exhibit multifractality, (iii) illustrate how this multifractality can be stable in the presence of disorder away from these integrable limits, and (iv) show how the multifractality can be better understood by introducing a restricted version of the inverse participation ratio, restricted to $n$-photon sectors, which is in turn reflected in the dynamics of the photonic mode and is apparent in photon bunching and the two-photon blockade.

These different results indicate that, despite the presence of the disorder, the Hamiltonian can be effectively described in terms of collective spin operators. Such collective spin dynamics is expected for identical spins and naturally appears in models with sufficiently long-range interactions~\cite{mori_prethermalization_2019,defenu_out--equilibrium_2023,defenu_long-range_2023,lerose_theory_2023}. While the spins in our setup are not identical due to the disorder, we showed for the new integrable limit that they can be treated as such up until relaxation time scales scaling as $\sqrt{N}$, with $N$ the number of spin modes, provided that the energy is bounded in the limit $N \to \infty$. Remarkably, the numerical results on the dynamics indicate that such a description remains accurate even in the presence of disordered bare energies.
It would be interesting to further investigate when collective spin descriptions hold in the presence of disorder for cavity models and clarify the role of integrability, following similar studies for systems with long-range interactions \cite{lerose_theory_2023}.

The accessibility of exact eigenstates and eigenenergies has led to various studies of the dynamics in Richardson-Gaudin models~\cite{bortz_exact_2007,faribault_quantum_2009,foster_quantum_2013,faribault_integrability-based_2013,claeys_spin_2018,villazon_shortcuts_2021,marino_dynamical_2022}, and this work opens up avenues to further study the dynamics in disordered Tavis-Cummings models.
A natural extension of this work is to consider open systems. Here we only note that the model with inhomogeneous interaction strengths remains solvable if we choose $\Delta$ to be complex, leading to non-Hermitian and hence dissipative dynamics, since the calculation of the eigenstates did not depend on $\Delta$ being real. The dynamics generated by a non-Hermitian Richardson-Gaudin Hamiltonian can be directly studied, as e.g. done in Refs.~\cite{rowlands_noisy_2018,claeys_dissipative_2022}, with only minimal modifications of the presented framework. An additional extension is to consider the model with an arbitrary number of excitations, where it is expected that many of the results presented in this work hold more generally.

\section*{Acknowledgements} 
We thank Stijn De Baerdemacker for useful comments and for clarifying the connection with the random phase approximation and Gabriel O. Alves for useful comments on the manuscript. This research was supported in part by the National Science Foundation under Grants No. NSF PHY-1748958 and PHY-2309135 at the Kavli Institute for Theoretical Physics (KITP).

\section{One-excitation Bethe ansatz}
\label{app:Bethe_derivation}

In this Appendix we briefly review some aspects of integrable factorizable Richardson-Gaudin Hamiltonians~\cite{claeys_richardson-gaudin_2018,rombouts_quantum_2010,Lukyanenko2016,bogolyubov_algebraic_2000,Skrypnyk2006,Skrypnyk2007,Skrypnyk2009a,Skrypnyk2009b,Lukyanenko2014,Lukyanenko2016,iyoda_effective_2018,
dukelsky_class_2001,dukelsky_colloquium_2004, ortiz_exactly-solvable_2005,rombouts_quantum_2010,van_raemdonck_exact_2014}. A more complete overview can be found in Sec.~3 of Ref.~\cite{tang_integrability_2023}. The family of the integrable factorizable Hamiltonians can be written as:
\begin{align}
    \label{eq:factorizable}
    \hat{H}(\alpha) &= \frac{1+\alpha}{2}\G^+\G^- + \frac{1-\alpha}{2}\G^-\G^+ \nonumber \\
    &=\alpha \sum_{i=1}^N g_i^2 \S_i^z + \frac{1}{2}\sum_{i,j=1}^N g_i g_j(\S_i^+\S_j^- + \S_i^-\S_j^+)    
\end{align}
where $\G^\pm = \sum_{i = 1}^{N}g_j \S_j^\pm$. 
We will be mostly interested in the solutions within the spin-$1/2$, one-excitation sector (i.e. the Hilbert space spanned by the basis $\S_i^\dagger |\emptyset \rangle$, $i=1\dots N$). Within this sector, the eigenstates can be written as Bethe states:
\begin{equation}
    \label{eq:singlebethe}
    \ket{\nu} = \G^+(\nu) |\emptyset\rangle, \quad \text{with} \quad \G^+(\nu) = \sum_{i = 1}^N \frac{g_i}{g_i^2 - \nu} \S^+_i,
\end{equation}
with eigenvalue,
\begin{equation}
    \label{eq:betheeigenvalue}
    (\alpha -1)\left[ \nu - \frac{1}{2}\sum^N_{i = 1}g_i^2\right],
\end{equation}
provided the Bethe equation
\begin{equation}
    \label{eq:betheequation}
    \frac{\alpha -1}{2} - \frac{1}{2}\sum_{j = 1}^N \frac{g_j^2}{g_j^2 - \nu} = 0,
\end{equation}
is satisfied. The derivation of this result is particularly simple in the one excitation case and included here in order to be self-contained.

Applying the Hamiltonian~\eqref{eq:factorizable} to the state \eqref{eq:singlebethe} and using the fact that $G^- \ket{0}= 0$ gives:
\begin{equation}
    \label{eq:Hcommutator}
    \hat{H}(\alpha)\ket{\nu}= [\hat{H}(\alpha),\G^+(\nu)] \ket{\emptyset} + \frac{(1-\alpha)}{2} \sum_{i = 1}^N g_i^2 \ket{\nu}.
\end{equation}
We will now focus on finding the commutator present above and begin by using the $SU(2)$ commutation relations of the spin operators to calculate:
\begin{equation}
    [\G^+, \G^+(\nu)] = 0, \quad [\G^-, \G^+(\nu)] = -2\sum_{i = 1}^N \frac{g_i^2}{g_i^2 - \nu} \S_i^z.
\end{equation}
The commutator in \eqref{eq:Hcommutator} can now be rewritten as
\begin{align}
    [\hat{H}(\alpha),\G^+(\nu)] &= -\sum_{i = 1}^N \frac{g_i^2}{g_i^2-\nu} \left(2\G^+ \S_i^z + (1-\alpha)[\S_i^z,\G^+]\right) \nonumber \\
    &= -\sum_{i = 1}^N \frac{g_i^2}{g_i^2-\nu} \left(2\G^+ \S_i^z + (1-\alpha)g_i \S_i^+\right)\,.
\end{align}
Applying this result to the vacuum state, substituting the definition of $\G^+$, and relabeling dummy variables in summations, we end up with the result
\begin{align}
    &[\hat{H}(\alpha),\G^+(\nu)] \ket{\emptyset} =(\alpha - 1)\frac{\nu g_i}{g_i^2 - \nu}S^+_i \ket{0}  \nonumber \\
    &\quad+\sum_{i = 1}^N g_i \left( - \sum_{j = 1}^N \frac{g_j^2}{g_j^2- \nu} + (\alpha - 1) \right) \S^+_i \ket{0}\,.
\end{align}
Using the definition of $| \nu \rangle$, our initial equation \eqref{eq:Hcommutator} becomes
\begin{align}
    \hat{H}(\alpha)|\nu \rangle &= (\alpha - 1)\left(\nu - \frac{1}{2}\sum_{i = 1}^N g_i^2 \right)\ket{\nu} \nonumber \\ 
    &+\sum_{i = 1}^N g_i \left( - \sum_{j = 1}^N \frac{g_j^2}{g_j^2- \nu} 
    + (\alpha - 1) \right) \S^+_i |0 \rangle \,.
\end{align}
In order  for $\ket{\nu}$ to be an eigenstate of the Hamiltonian, we require $\hat{H}(\alpha) \ket{\nu} = E\ket{\nu}$, where $E$ is the eigenvalue of the eigenstates. Imposing this on the above expression, we obtain the correct Bethe equation along with the expected eigenvalue for our initial state.

\section{Derivation of the conserved charges}
\label{app:charges}
In this Appendix we explicitly derive the commutation relations of the conserved charges. The derivation is analogous to a similar derivation for the conserved charges of a spin-1 central spin model as presented in Ref.~\cite{tang_integrability_2023}. These conserved charges are again constructed by using the properties of the integrable Richardson-Gaudin models \eqref{eq:factorizable}, where $\hat{H}(\alpha)$ has conserved charges
\begin{align}
\hat{Q}_j(\alpha)& = \alpha \S_j^z +\tilde{Q}_j,
\end{align}
with $\tilde{Q}_j$ defined in Eq.~\eqref{eq:factorizablecharges} in the main text. These charges satisfy
\begin{align}
[\hat{H}(\alpha),\hat{Q}_j(\alpha)]=[\hat{Q}_j(\alpha),\hat{Q}_k(\alpha)]=0,
\end{align}
for all $j, k = 1, \ldots, L$. The conserved charges from the main text are given in block matrix representation by
\begin{align}
\hat{Q}_j = \begin{pmatrix}
-\frac{2}{3} \Delta (\tilde{Q}_j + \S_j^z) & \hat{G}^+ (\tilde{Q}_j + \S_j^z/3) & 0 \\
(\tilde{Q}_j+\S_j^z/3)\hat{G}^- & \frac{\Delta}{3}(\tilde{Q}_j+3 \S_j^z) & \sqrt{2}(\tilde{Q}_j + \S_j^z/3)\hat{G}^+ \\
0 & \sqrt{2}\hat{G}^-(\tilde{Q}_j + \S_j^z/3) & \frac{4}{3}\Delta(\tilde{Q}_j - \S_j^z).
\end{pmatrix}\nonumber
\end{align}
where the different blocks correspond to different photon number. The Tavis-Cummings Hamiltonian can similarly be represented as a block-diagonal matrix
\begin{align}
\hat{H} = \begin{pmatrix}
0 & \G^+ & 0 \\
\G^- & \Delta & \sqrt{2} \G^+ \\
0 & \sqrt{2}\G^- & 2 \Delta
\end{pmatrix}.
\end{align}
Using only these block matrix representations, the commutator of the Hamiltonian with these charges can be evaluated as 
\begin{widetext}
\begin{align}
&[\hat{H},\hat{Q}_j] = \nonumber\\
&\begin{pmatrix}
0 & \frac{2}{3}\Delta ( \G^+ (-\tilde{Q}_j+\S_j^z)+(\tilde{Q}_j+\S_j^z)\hat{G}^+) & 0 \\
\frac{2}{3}\Delta ((-\tilde{Q}_j+\S_j^z)\G^- +(\tilde{Q}_j+\S_j^z)\G^- ) & [\tilde{Q}_j+\S_j^z/3,\G^-\G^+ + 2 \G^+ \G^-] & \frac{4\sqrt{2}}{3}\Delta ( \G^+ (\tilde{Q}_j-\S_j^z)-(\tilde{Q}_j+\S_j^z)\G^+)  \\
0 & \frac{4\sqrt{2}}{3}\Delta ( (\tilde{Q}_j-\S_j^z)\G^- -\G^-(\tilde{Q}_j+\S_j^z))  & 0
\end{pmatrix}
\end{align}
\end{widetext}
where we have not yet made use of any properties of the operators. The commutator on the diagonal vanishes since $\tilde{Q}_j+\S_j^z/3 = \hat{Q}_j(\alpha=1/3)$ is exactly a conserved charge for $\hat{H}(\alpha=1/3)\propto \G^-\G^+ + 2 \G^+ \G^-$. The off-diagonal elements vanish because of the identity
\begin{align}
\G^+ (\tilde{Q}_j-\S_j^z) =  (\tilde{Q}_j+\S_j^z)\G^+,
\end{align}
which can be checked either from direct calculation or by noting that $\G^+ \hat{H}(\alpha=-1) = \hat{H}(\alpha=1)\G^+$. This identity relates the eigenstates of $\hat{H}(\alpha=-1)$ and $\hat{H}(\alpha=1)$, as also discussed in Refs.~\cite{villazon_integrability_2020,tang_integrability_2023}, and since the conserved charges share a common set of eigenstates with these Hamiltonians this identity should also hold on the level of the conserved charges, and we can rewrite the above equation as
\begin{align}
\G^+ \hat{Q}_j(\alpha=-1) = \hat{Q}_j(\alpha=1)\G^+\,.
\end{align}
Since all matrix elements of the commutator vanish in the block matrix representation, we hence find that the Hamiltonian commutes with all proposed conserved charges.

\section{Secular equation with vanishing poles}
\label{app:vanishing_pole}
\begin{figure}[t!]                      
 \begin{center}
 \includegraphics[width=\columnwidth]{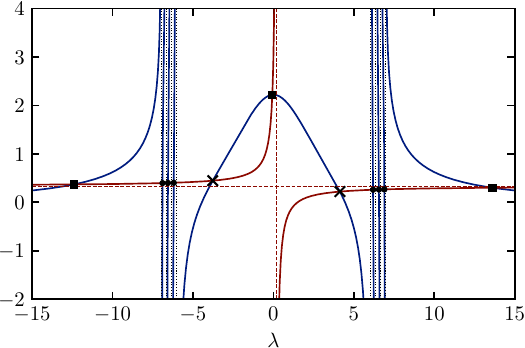}
 \caption{Graphical illustration of the secular equation in the case where a pair of poles vanishes. Each intersection between the left-hand side (red line) and right-hand side (blue line) returns the eigenvalue of a polariton state, leading to different classes of solutions. Parameters: $\Delta=1$, $\epsilon=0$, $N=5$ and $g_i^2 \in [2,4,6,8,32]$.
 \label{fig:secular_pole}}
 \end{center}
\vspace{-\baselineskip}
\end{figure}
The pole structure of the secular equation~\eqref{eq:secular},
\begin{equation}\label{eq:app_secular}
    \frac{\lambda - 3\tilde{\Delta}}{3\lambda - \tilde{\Delta}} = \sum_{i = 1}^{N}\frac{g_i^2}{\lambda^2 - \tilde{\Delta}^2 - \bar{g}^2 + 2g_i^2}\,,
\end{equation}
can change abruptly if the inhomogeneous interaction strengths are strongly asymmetrically distributed. 
Here $\bar{g}^2 = \sum_{i = 1}^N g_i^2$ and $\tilde{\Delta} = \Delta/2$. It is now possible for a pair of poles to vanish whenever
\begin{align}
2 g_i^2-\bar{g}^2-\tilde{\Delta}^2 \geq 0,
\end{align}
which can be rewritten as
\begin{align}
g_i^2 \geq \tilde{\Delta}^2 + \sum_{j \neq i}^N g_j^2\,.
\end{align}
This situation occurs whenever a single $g_i$ is sufficiently large compared to the remaining interaction strengths. This equation can clearly only be satisfied for a single $g_i$, such that, in addition to the discussion of the main text, the only additional case that needs to be considered is the one where a single pair of poles vanishes.

Both sides of the resulting secular equation are plotted in Fig.~\ref{fig:secular_pole}. There are again $2N+1$ intersections between both sides, indicating $2N+1$ solutions, such that we obtain the corect number of states. While the total number of solutions remains the same, the direct identification between solutions to the secular equation and eigenstates in the homogeneous limit now breaks down. The pole structure results in $2N-4$ solutions in between pairs of poles (marked by circles), and there are $5$ additional solutions. The triplet polariton solutions can again be identified near the vertical asymptote of the left-hand side and for large $\lambda$ (as marked by squares), but two additional solutions now appear in the middle interval (marked by crosses). These additional solutions always appear, since the right-hand side of Eq.~\eqref{eq:app_secular} lies above the horizontal asymptote of the right-hand side for $\lambda$ around zero.

Let us focus on the center interval where the two additional poles appear. The number of intersection will be determined by the behavior of the right-hand side for $\lambda \to 0$. In the case from the main text no poles vanish and it directly follows that for $\lambda \to 0$ the right-hand side will always be negative and hence below the horizontal asymptote $1/3$ of the left-hand side, such that there are no additional intersections with the left-hand side. If a pole vanishes, then for $\lambda \to 0$ the right-hand side can be shown to be larger than $1$ and hence above the horizontal asymptote $1/3$, introducing two additional intersections.

That the right-hand side is always larger than 1 for $\lambda=0$ can be directly checked. Assuming that the pole corresponding to $g_1$ vanishes, the right-hand side for $\lambda=0$ can be written as
\begin{align}
\frac{g_1^2}{g_1^2-\rho^2} - \sum_{i = 2}^{N}\frac{g_i^2}{g_i^2-g_1^2-\rho^2}\,
\end{align}
where $\rho^2 = \tilde{\Delta}^2 + \sum_{j = 2}^N g_j^2 $ such that $g_1^2 \geq \rho^2$ (in order for the pole to vanish) and $g_1^2 \geq g_i^2$. For $g_1^2 \to \infty$ this expression approaches $1$, for $g_1^2 \to \rho^2$ from above this expression diverges to $+\infty$. The derivative w.r.t. $g_1^2$ can easily be checked to be negative for all values of $g_1^2$ in between these two limits, such that this expression is monotonous between these limits.

\section{IPR scalings from the Bethe ansatz}
\label{app:IPR_from_Bethe}

The multifractality scalings can be directly obtained from the Bethe state \eqref{eq:BetheAnsatz_2} and reflect the different positions of the solutions w.r.t. the poles. From Eq.~\eqref{eq:BetheAnsatz_2} we have that
\begin{align}
&\tilde{\psi}_0 = \sqrt{2}, \\
&\tilde{\psi}_i = -\frac{g}{\epsilon_i-E_1}-\frac{g}{\epsilon_i-E_2}, \\
&\tilde{\psi}_{i,j} = \frac{g^2}{(\epsilon_i-E_1)(\epsilon_j-E_2)}+\frac{g^2}{(\epsilon_i-E_2)(\epsilon_j-E_1)},
\end{align}
where $\tilde{\psi}$ equals ${\psi}$ up to a global normalization factor, since the Bethe states are unnormalized.

We can now identify different scaling behaviors depending on the relative positions of the variables $E_{1,2}$ w.r.t. the poles $\epsilon_i,i=1\dots N$, in the same way that the relative position of the root to the secular equation \eqref{eq:secular} w.r.t. the poles $g_i, i=1\dots N$ allowed us to distinguish different classes of eigenstates. The derivations for both integrable limits are highly similar, but since the IPR cannot be defined for the degenerate dark states in the limit of homogeneous bare energies, we first focus on the limit of homogeneous interaction strength and inhomogeneous bare energies.

\emph{Triplet polaritons.} Consider a situation where the two Bethe roots $E_{1,2}$ are a distance $O(1)$ away from all poles $\epsilon_i, i=1 \dots N$. In this scenario all terms $1/(\epsilon_i-E_{\alpha})$ will be $O(1)$, such that the scaling of the amplitudes is purely set by the scaling of the factor $g= O( N^{-1/2})$. In this scenario, we find that
\begin{align}
|\tilde{\psi}_0|^{2q} = O(1),\quad |\tilde{\psi}_i|^{2q} = O(N^{-q}),\quad |\tilde{\psi}_{i,j}|^2 = O(N^{-2q}), \nonumber
\end{align}
which holds $\forall i,j$, such that summing over the appropriate number of terms in the restricted IPRs returns the obtained (ergodic) scalings
\begin{align}
\mathcal{P}_0(q) &\propto |\tilde{\psi_0}|^{2q} = O(1), \nonumber\\
\mathcal{P}_1(q) &\propto \sum_i |\tilde{\psi}_i|^{2q}  = O(N) \times O(N^{-q}) =  O(N^{1-q}), \nonumber\\
\mathcal{P}_2(q)&\propto \sum_{i,j} |\tilde{\psi}_{i,j}|^{2q} = O(N^2) \times O(N^{-2q}) = O(N^{2-2q})\,. \nonumber
\end{align}
Each first term in a product is the number of components and the second term is the scaling of the individual components. 
Crucially, we find that $\mathcal{P}_0(q=1) = \mathcal{P}_1(q=1) = \mathcal{P}_2(q=1) = O(1)$, such that $\tilde{\psi}$ and $\psi$ have the same scaling. These results then reproduce the observed scalings from the main text.

\begin{figure}[t!]                      
 \begin{center}
 \includegraphics[width=\columnwidth]{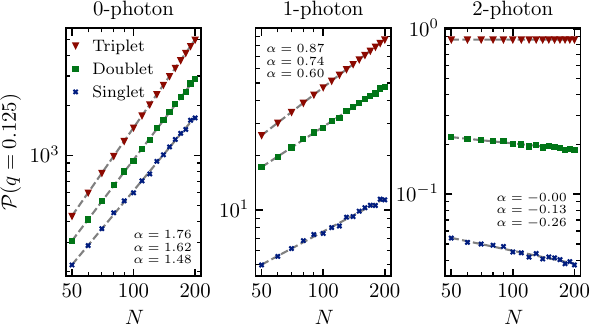}
 \caption{Scaling of the IPR for the three different classes of eigenstates in the different subsection of the Hilbert space for $q=0.125$ and an integrable model with disordered bare energies. Gray dashed lines indicate best fits $\propto N^{\alpha}$. Parameters: $\Delta = 1$, $g=2/\sqrt{N}$ and $\epsilon_i$ uniformly distributed in $[-0.1,0.1]$.
 \label{fig:IPR:homo_g}}
 \end{center}
\vspace{-\baselineskip}
\end{figure}
\begin{figure}[t!]                      
 \begin{center}
 \includegraphics[width=\columnwidth]{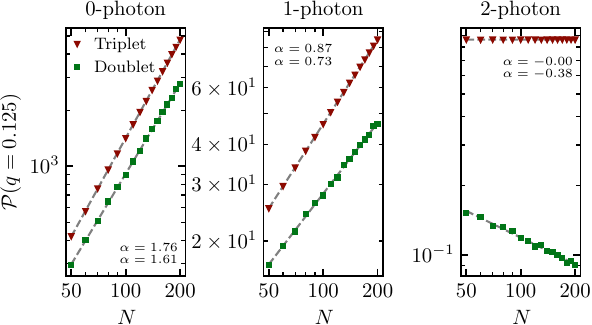}
 \caption{Scaling of the IPR for the polaritons in the different subsections of the Hilbert space for $q=0.125$ and an integrable model with disordered interaction strengths. Gray dashed lines indicate best fits $\propto N^{\alpha}$. Parameters: $\Delta = 1$,  $\epsilon_i=0$ and $g_i$ uniformly distributed in $[1,3]/\sqrt{N}$.
 \label{fig:IPR:homo_eps}}
 \end{center}
\vspace{-\baselineskip}
\end{figure}
\emph{Doublet polaritons.} In this case a single Bethe root, e.g. $E_1$, lies in between two poles, where the typical distance between two poles scales as $O(N^{-1})$, and the other pole is again a distance $O(1)$ away from all poles. The summations appearing in the (restricted) IPR now needs to be separated in the contributions close to the poles, where $1/(\epsilon_i-E_{\alpha}) = O(N)$, and the contributions further from the poles, following a similar argument in Ref.~\cite{dubail_large_2022}. 

For $|\tilde{\psi}_i|^2$, an $O(1)$ number of components will have contributions $O(N)$ due to the proximity of the Bethe root to the pole, whereas the remaining $O(N)$ terms can again be treated as contributing a scaling $O(N^{-1})$. In the same way, the 0-photon terms $|\tilde{\psi}_{i,j}|$ will have $O(N)$ terms where $E_1$ is close to the pole $\epsilon_i$ or $\epsilon_j$, and these terms individually scale as $O(1)$, whereas the remaining $O(N^2)$ terms scale again scale as $O(N^{-2})$. The terms scaling as $O(N)$ in $|\tilde{\psi}_i|^2$ will dominate the norm, combined with the $O(N)$ terms scaling as $O(1)$ in $|\tilde{\psi}_{i,j}|^2$, such that the total norm scales as $O(N)$ and $|\psi|^2 = O(N^{-1})|\tilde{\psi}|^2$. Rescaling all components by the appropriate normalization factor, the restricted IPRs follow as
\begin{align}
\mathcal{P}_0(q) &= 1 \times O(N^{-q}) = O(N^{-q}), \\
\mathcal{P}_1(q) &= O(1) \times O(1) + O(N) \times O(N^{-2q}) \nonumber\\
&=  O(1)+O(N^{1-2q}),\\
\mathcal{P}_2(q) &= O(N) \times O(N^{-q}) + O(N^2) \times O(N^{-3q}) \nonumber\\
&=  O(N^{1-q})+O(N^{2-3q}),
\end{align}
where each first term in a product is again the number of components and the second term is the scaling of the individual components. These results reproduce the observed scalings from the main text.

\emph{Singlet dark states.} For the dark states both Bethe roots are close to a pole. The crucial difference with the double polaritons is now that there are components $\tilde{\psi}_{i,j}$ where e.g. $\epsilon_i$ is close to $E_1$ and $\epsilon_j$ is close to $E_2$, such that these components will dominate the wave function (and give rise to semilocalization). In this case $|\tilde{\psi}{i,j}|^2 = O(N^2)$. There will be $O(N)$ remaining components $|\tilde{\psi}_{i,j}| = O(1)$, where a single Bethe root is close to a pole, and $O(N^2)$ components $|\tilde{\psi}_{i,j}| = O(N^{-2})$, where no Bethe root is close to a pole. The remaining components behave identical to the doublet polariton case: there are $O(1)$ terms $|\tilde{\psi}_i|^2 = O(N)$ and $O(N)$ terms $|\tilde{\psi}_i|^2 = O(N^{-1})$, and $|\tilde{\psi}_0|^2 = O(1)$. Crucially, the norm now scales different because of the additional terms in $|\tilde{\psi}_{i,j}|^2$, and $|\tilde{\psi}|^2 = O(N^{-2})|\psi|^2$. 

Introducing this rescaling, the resulting restricted IPRs follow as
\begin{align}
\mathcal{P}_0(q) &= 1 \times O(N^{-2q}) = O(N^{-2q}) \\
\mathcal{P}_1(q) &= O(1) \times O(N^{-q})+O(N)\times O(N^{-3q}) \nonumber\\
& = O(N^{-q}) + O(N^{1-3q}) \\
\mathcal{P}_0(q) &= O(1) \times O(1)+O(N) \times O(N^{-2q}) \nonumber\\
&\qquad +O(N^2)\times O(N^{-4q}) \nonumber\\
&= O(1) + O(N^{1-2q})+O(N^{2-4q})\nonumber\\
&= O(1) +O(N^{2-4q}),
\end{align}
returning the scaling from the main text. 

The numerically obtained IPRs in this integrable limit are illustrated in Fig.~\ref{fig:IPR:homo_g} for all classes of eigenstates, and are visually indistinguishable from the result in presence of disordered interaction strengths.

These results can be contrasted with the IPR scaling in the opposite integrable limit, where the bare energies are homogeneous and the interaction strengths are disordered. Since the dark singlet states are exactly degenerate any linear combination of dark states would return a dark state, such that it is not meaningful to consider the localization properties of dark states. We will only focus on the IPR for the triplet and doublet polaritons. The scaling again follows from the relative position of the roots to the poles of the secular equation~\eqref{eq:secular}. The components of the unnormalized eigenstates follow from Eq.~\eqref{eq:bright_ansatz} as
\begin{align}
\tilde{\psi}_0 &= \frac{1}{\kappa-\Delta}\sum_{i=1}^N \frac{g_i^2}{\lambda^2 -  \tilde{\Delta}^2 -\bar{g}^2 + 2g_i^2} = \frac{1}{3 \lambda-\tilde{\Delta}}\,, \\
\tilde{\psi}_i &=\frac{g_i}{\lambda^2 -  \tilde{\Delta}^2 -\bar{g}^2 + 2g_i^2}\,, \\
\tilde{\psi}_{i,j} &=\frac{1}{\kappa+\Delta}\frac{g_i g_j}{\lambda^2 -  \tilde{\Delta}^2 -\bar{g}^2 + 2g_i^2} + (i\leftrightarrow j)\,.
\end{align}
In $\tilde{\psi}_0$ we have used the secular equation~\eqref{eq:secular}. For the triplet polaritons $\lambda$ is $O(1)$ removed from the poles, and we find that $|\tilde{\psi}_0|^2 = O(1)$, $|\tilde{\psi}_i|^2 = O(N^{-1})$ and  $|\tilde{\psi}_{i,j}|^2 = O(N^{-2})$ due to the scaling of $g_i^2 = O(N^{-1})$. These scalings are identical to the opposite integrable limit and hence result in the same expression for the restricted IPRs:
\begin{align}
\mathcal{P}_0(q) & = O(1), \nonumber\\
\mathcal{P}_1(q) & = O(N) \times O(N^{-q}) =  O(N^{1-q}), \nonumber\\
\mathcal{P}_2(q)& = O(N^2) \times O(N^{-2q}) = O(N^{2-2q})\,. \nonumber
\end{align}
For the doublet polariton the root is close to the poles, and the distance to the closest pole is again on the order of the spacing between two nearest poles, which now however scales as $O(N^{-2})$. Plugging in this scaling, the number $O(1)$ components $|\tilde{\psi}_i|^2$ close to the pole scale as $O(N^3)$ and the remaining $O(N)$ components scale as $O(N)$. Similarly, for $|\tilde{\psi}_{i,j}|^2$ there are $O(N)$ components close to the pole scaling as $O(N^2)$, and the remaining $O(N^2)$ components are $O(1)$. The total norm hence scales as $O(N^3)$, and a rescaling of these components by these factors results in a restricted IPR
\begin{align}
\mathcal{P}_0(q) &= 1 \times O(N^{-3q}) = O(N^{-3q}) \\
\mathcal{P}_1(q) &= O(1) \times O(1) + O(N) \times O(N^{-2q}) \nonumber\\
&=  O(1)+O(N^{1-2q}),\\
\mathcal{P}_2(q) &= O(N) \times O(N^{-q}) + O(N^2) \times O(N^{-3q}) \nonumber\\
&=  O(N^{1-q})+O(N^{2-3q}),
\end{align}
The numerically obtained IPR is shown in Fig.~\ref{fig:IPR:homo_eps} for the doublet and triplet polaritons, and it is clear that only $\mathcal{P}_0(q)$ differs, scaling as $N^{-3q}$ as opposed to $N^{-q}$ in the other discussed cases.

\bibliography{TavisCummingsLibrary}

\end{document}